\documentstyle[aps]{revtex}
\input epsf
\title{\bf Correlation functions in the factorization approach
of nonextensive quantum statistics}
\author {Marcelo R. Ubriaco\thanks{Electronic address:ubriaco@ltp.upr.clu.edu}}
\address {Laboratory of Theoretical Physics\\
 Department of Physics\\
 University of Puerto Rico\\
 P. O. Box 23343, R\'{\i}o Piedras\\
 PR 00931-3343, USA}
\begin{document} 
\vspace{0.3in}
\maketitle
\vspace{0.15in}
\begin{abstract}
We study the long range behavior
of a gas whose partition function depends on
a parameter $q$ and it has been claimed
to be a good approximation to the partition function
proposed in the formulation of
nonextensive statistical mechanics. We compare our results,
at large temperatures and at the critical point, with the
 case of Boltzmann-Gibbs thermodynamics for the
case of a Bose-Einstein gas. In particular, we find that
for all temperatures the long range correlations in a Bose gas decrease when
the value of $q$ departs from the standard value
$q=1$.
\end{abstract}
\vspace*{.2in}
PACS number(s):05.30.-d
\baselineskip20pt
\section{Introduction}
The formulation of nonextensive statistical mechanics \cite{T1} has raised numerous
questions regarding its relevance to systems with long range interactions, long range
microscopic memory, or multifractal properties  \cite{T2}.  This formalism
can  also be understood as a generalization of Boltzmann-Gibbs statistics,
opening  the possibility to get a theoretical insight on the
  thermodynamics of systems whose behavior depart
from Boltzmann-Gibbs statistical mechanics.
The generalized entropy 
\begin{equation}
S_q=\frac{k}{q-1}\left(1-\sum_Rp^q_R\right),\label{Sq}
\end{equation}
is a function of the probability $p_R$ for the ensemble to be in the state $R$
and a real parameter $q$.  Equation (\ref{Sq})
becomes the Shannon entropy as $q\rightarrow 1$. In addition,  Tsallis'
entropy shares all the properties of the Shannon entropy except that of additivity.
Considering two independent systems $\Sigma$ and $\Sigma'$, the entropy $S_q$ satisfies
the pseudoadditivity property
\begin{equation}
\frac{S_q^{\Sigma\cup\Sigma'}}{k}=\frac{S_q^{\Sigma}}{k}+
\frac{S_q^{\Sigma'}}{k}+(1-q) \frac{S_q^{\Sigma}}{k}\frac{S_q^{\Sigma'}}{k} .
\end{equation}
The  probability distribution that results from extremizing the entropy with the
constraints 
\begin{equation}
\sum_Rp_R=1,
\end{equation}
 and 
\begin{equation}
\langle E\rangle=\sum_Rp^q_RE_R,\label{E}
\end{equation} 
is given by the equation
\begin{equation}
p_R=\frac{\left[1+\beta(q-1)(E_R-\mu N)\right]^{1/(1-q)}}{Z_q},
\end{equation}
with the partition function
\begin{equation}
 Z_q=\sum_R\left[1+\beta(q-1)(E_R-\mu N)\right]^{1/(1-q)} \label{Zq},
\end{equation}
and the total energy $E_R=\sum_j n_j\varepsilon_j$.
Due to the mathematical complexity of the partition function 
in Equation (\ref{Zq}), it has been necessary to study the validity of
certain  approximation procedures.  Thus,  the study of the
consequences of nonextensivity has been mainly focused on either assuming a value of $q\approx 1$ 
\cite{T3}  or by approximating the partition function in Eq. (\ref{Zq}) by a factorized partition function 
\cite{BDG}
\begin{equation}
Z=\prod_{j=0}\sum_{n_j=0}\left[1+\beta(q-1)n_j(\varepsilon_j-\mu)\right]^{1/(q-1)}.
\end{equation}
The factorization approach has been shown \cite{WM} to be a good approximation
to Tsallis partition function outside certain narrow temperature interval that shifts to
higher values of $T$ when the number of energy levels increases. An application to 
the Ising model and black body radiation can be found in Refs. \cite{BDT} and \cite{WM2}
respectively.
In this approximation the average number of particles with energy $\varepsilon$
is given by the function
\begin{equation}
\langle n\rangle=\frac{1}{\left[1+\beta (q-1)(\varepsilon-\mu)\right]^{1/(q-1)}+a},\label{n}
\end{equation}
where $a=0,-1,+1$ for Maxwell-Boltzmann, Bose-Einstein and Fermi-Dirac cases respectively.
It is important to remark that the function $\langle n\rangle$ has also been obtained
\cite{BD} by extremizing the entropy related to the generalized dimensions of 
a fractal set.
It has been also pointed out \cite{T4} that nonextensive thermodynamics could also be  
understood in terms of $q$-deformations and possibly with the theory of
quantum groups.  
Along this line of work, in Ref. \cite{U1} we made a study of the basic thermodynamics 
that result from the particle distribution functions
in Eq. (\ref{n}) for classical and quantum gases.  In that article we shown 
that the high temperature behavior is consistent with the thermodynamical limit 
provided that the internal energy is calculated according to the equation
\begin{equation}
\langle U\rangle=\frac{4\pi V}{h^3}\int_0^{\infty}\frac{p^2}{2m}\langle n(p)\rangle^qp^2 dp.
\end{equation}
 By introducing  the usual creation and annihilation operator formalism we 
found that the boson hamiltonian that leads to the particle distribution, for $a=-1$, in 
Eq.(\ref{n}) is written as
\begin{equation}
 \hat{K}=\sum_{j=0}(\varepsilon_j-\mu)\bar{\phi}_j\phi_j,
\end{equation}
where the operator $\phi_j$ and its adjoint $\hat{\phi}_j$ satisfy a deformed
boson algebra.
However, a comparison  of the heat capacity and entropy functions, in Ref. \cite{U1},
for systems with  a particle distribution function as given by Eq. (\ref{n}) with
those for Bose and Fermi gases described by quantum group invariant
hamiltonians \cite{U2} shows that nonextensivity is unrelated to quantum group invariance.

In this paper we calculate the correlation function for boson systems with a 
particle distribution function as given by Eq. (\ref{n}).  Our main motivation
in studying  this system is twofold.  First, it is of theoretical interest to
study a thermodynamics system which obeys a statistical mechanics that 
generalizes Boltzmann-Gibbs statistics.  Second, a calculation  of
the correlation functions will give us an insight on the
long-range behavior dependence of these thermodynamical systems 
on the parameter $q$.  Our calculations will show in a concrete
fashion  the
relation between $q$ and long-range behavior, and will indicate
whether the thermodynamics resulting from Eq. (\ref{n}),
proposed in Ref.\cite{BDG}, has the long range behavior
expected to be present in nonextensive thermodynamics.
In Sec. \ref{2} we calculate the correlation function for $q\neq 1$, and discuss 
the results for some particular values of this parameter. 
We specialize our discussion to the behavior near the critical
temperature $T_c$ and at  large values of $T$.  In Sec. \ref{3}  we summarize our results.
\section{ Correlation functions } \label{2}
According to our previous work \cite{U1},
the parameter $q$ cannot be any real number 
but its values are restricted to those such that $1/(q-1)$ is an integer.  In addition,
the thermodynamic functions are well defined only in the interval $1\leq q\leq 1.5$.
The correlation for a Bose gas with particle distribution  according to Eq.(\ref{n}) 
 is given by
\begin{equation}
G({\bf R})=\frac{1}{V}\sum_{{\bf k}}\frac{e^{i{\bf k\cdot R}}}
{\left[1+\beta (q-1)(\varepsilon-\mu)\right]^{1/(q-1)}-1},
\end{equation}
As usual, this summation can be approximated by an integration over ${\bf k}$,
leading after integration over the angles for $d=3$ to the equation 
\begin{equation}
G({\bf R})=\frac{\langle N_0\rangle}{V}+\frac{2}{\pi^{1/4}R^{1/2}(\lambda\sqrt{q-1})^{5/2}}
\int_0^{\infty}x^{3/2}\sum_{j=1}^{\infty}\left[1+\alpha(q-1)+x^2\right]^{-j/(q-1)}
J_{1/2}\left(2\pi^{1/2}Rx/\lambda\sqrt{q-1}\right)dx,\label{G}
\end{equation}
where the new variable $x^2=\beta(q-1)k^2\hbar^2/2m$ and, as usual, $\alpha=-\beta\mu$
and $\lambda$ is the thermal wavelength.  
The integral in Eq. (\ref{G}) is tabulated \cite{GR}
\begin{equation}
\int_0^{\infty}\frac{J_{\nu}(bx)}{(x^2+c^2)^{\mu +1}}x^{\nu +1}dx=\frac{c^{\nu-\mu}b^{\mu}}
{2^{\mu}\Gamma(\mu +1)}K_{\nu -\mu}(cb),
\end{equation}
where $-1<Re \;\nu<Re(2\mu+3/2)$, $c>0$ and $b>0$.  It is simple to check that all 
these inequalities are satisfied for all the allowed values of $q$ in the interval $1\leq q\leq 1.5$.
Defining a length parameter $\chi<\lambda$
\begin{equation}
\chi=\frac{\lambda}{2\sqrt{\pi}}\sqrt{\frac{q-1}{1+\alpha (q-1)}},
\end{equation}
and by use of the integral representation 
\begin{equation}
K_{\nu}(zx)=\frac{\Gamma(\nu+1/2)}{x^{\nu}\Gamma(1/2)}(2z)^{\nu}\int_O^{\infty}\frac{\cos(xt)}{(t^2+z^2)^{\nu+1/2}}dt,
\end{equation}
the correlation function reduces to
the expression
\begin{equation}
G({\bf R})= \frac{1}{\pi R\lambda^2(q-1)}\int_0^{\infty}\cos(t)\sum_{n=1}\left[\frac{(R/\chi)^2}
{(1+\alpha(q-1))(t^2+(R/\chi)^2)}
\right]^{n/(q-1)-1} dt.
\end{equation}
Once we perform the summation we get a set of integrals in terms of elementary functions. These
integrals are all tabulated \cite{GR1} and the correlation function for some values of $q$ follows
\\
\begin{equation}
G({\bf R})=\left\{
\begin{array}{lll}
(1/ R\lambda^2)\left[e^{-R/\xi}-e^{-\sqrt{(R/\xi)^2+32\pi (R/\lambda)^2}}-2e^{-A'}\sin(B')\right] & \mbox{for $q=5/4$}\\ \\
(1/R\lambda^2) \left[e^{-R/\xi}-e^{-A}\cos(B)-\sqrt{3}e^{-A}\sin(B)\right] & \mbox{for $q=4/3$}\\ \\
(1/R\lambda^2)\left[e^{-R/\xi}-e^{-\sqrt{(R/\xi)^2+16\pi (R/\lambda)^2}}\right]&\mbox{for $q=3/2$}
\end{array}
\right.,
\end{equation}
\\
where $\xi=\lambda/2\sqrt{\pi\alpha}$ is the correlation length and the exponents
\begin{eqnarray}
A^2&=&12\pi(R/\lambda)^2\left[(1/2)\sqrt{(3/2+\alpha/3)^2+(3/4)}+(3/2+\alpha/3)\right]\nonumber\\
B^2&=&12\pi(R/\lambda)^2\left[(1/2)\sqrt{(3/2+\alpha/3)^2+(3/4)}-(3/2+\alpha/3)\right]\nonumber\\
A'^2&=&16\pi(R/\lambda)^2\left[(1/2)\sqrt{(1+\alpha/4)^2+1}+(1+\alpha/4)\right]\nonumber\\
B'^2&=&16\pi(R/\lambda)^2\left[(1/2)\sqrt{(1+\alpha/4)^2+1}-(1+\alpha/4)\right]\nonumber
\end{eqnarray}
From these equations it is clear that the correlation functions for $q\neq 1$ have a leading term
\begin{equation}
G({\bf R})\approx (1/R\lambda^2) e^{-R/\xi},\label{G2}
\end{equation}
minus some smaller additional terms which are absent in the Bose-Einstein case.
\subsection{Correlation length}
In order to compare these correlation functions with
the standard correlation function $G_{BE}({\bf R})$ we first look at
the case of high temperature $T>>T_c$. 
Although Eq. (\ref{G2}) has the same functional relation
than
\begin{equation}
G_{BE}({\bf R})\approx  (1/R\lambda^2) e^{-R/\xi_{BE}},
\end{equation}
 the parameter $\alpha$ is the function \cite{U1}
\begin{equation}
\alpha= \frac{1}{q-1}\left[-1+\left(\frac{-2}{\sqrt{\pi}\langle n\rangle(q-1)^{3/2}\lambda^3}S_2(q)\right)^{2(q-1)/(5-3q)}
\right],
\end{equation}
with
\begin{equation}
S_2(q)=\frac{1}{3/2-1/(q-1)}+\sum_{m=1}^{\infty}\frac{(-1)^m}{m!}(1/2)...(3/2-m)\frac{1}{3/2-m-1/(q-1)}.
\end{equation}
In terms of the critical temperature $T_c$ the  correlation length $\xi$ for high 
temperatures is given by the expression
\begin{equation}
\frac{1}{\xi}=2\sqrt{\frac{\pi}{q-1}}\left(\frac{\langle n\rangle}{G_{3/2}(1,q)}\right)^{1/3}
\left(\frac{T}{T_c}\right)^{1/2}\left[-1+\left(\frac{-2}{\sqrt{\pi}G_{3/2}(1,q)(q-1)^{3/2}}\left(\frac{T}{T_c}\right)^{3/2}
 S_2(q)\right)^{2(q-1)/(5-3q)}\right]^{1/2},
\end{equation}
where the function $G_{3/2}(z=1,q)$ was calculated in Ref. \cite{U1}.
In particular $G_{3/2}(1,1)=2.612$  and $G_{3/2}(1,q)$ increases with the value of $q$. A simple
inspection shows that the correlation length
$\xi$  is smaller than the correlation length for the
standard Bose-Einstein case $\xi^{BE}$
\begin{equation}
\frac{1}{\xi^{BE}}=2\sqrt{\pi}\left(\frac{\langle n\rangle}{2.612}\right)^{1/3}
 \left(\frac{T}{T_c^{BE}}\right)^{1/2}\ln^{1/2}\left[2.612(T/T_c^{BE})^{3/2}\right],
\end{equation}
where $T_c<T_c^{BE}$. Figure 1 shows a comparison between these two functions for several values
of the parameter $q$.  Clearly, the correlation function decreases more rapidly as $q$ increases
from the standard value $q=1$.  
At the critical temperature,  $\xi\rightarrow \infty$
and the 
correlations become
\begin{equation}
G_c({\bf R})\approx \frac{1}{R\lambda_c^2},
\end{equation}
which, due to the inequality $\lambda_c>\lambda_c^{BE}$, is smaller than
$G_c^{BE}$. Figure 2 shows  $\alpha$ for $T\geq T_{critical}$ and some values
of $q$ in comparison to the $q=1$ case. Since, at $T>T_c$,
$\alpha$ is smaller  for $q=1$ we find that the
correlation function for all temperatures is larger for the $q=1$ case.
\subsection{Critical behavior}
In order to study the behavior at the critical temperature we need to expand the function
\begin{equation}
G_{3/2}(z,q)=\frac{2}{\sqrt{\pi}(q-1)^{3/2}}\int_0^{\infty}\frac{y^{1/2}}{\left[1+y-(q-1)\ln z\right]^{1/(q-1)}-1}dy,
\end{equation}
in powers of $\alpha$.  It is clear from the previous discussion that $G_{3/2}(z,q)$ becomes
the well known Bose-Einstein function $g_{3/2}(z)$ in the limit $q\rightarrow 1$.
For these purposes we apply the same method used in Ref.\cite{R}, which gives a power
series of $\alpha$ by 
performing
first a Mellin transformation of the function and then applying the
corresponding inverse transform.   The Mellin transform
of $G_{3/2}(z,q)$ is given by the equation 
\begin{eqnarray}
F_{3/2}(s)&=&\int_0^{\infty}G_{3/2}(z,q) \alpha^{s-1}d\alpha, \nonumber\\
&=&\frac{\Gamma(s)}{(q-1)^{3/2+s}}\sum_{m=1}\frac{\Gamma\left(m/(q-1)-(3/2)-s\right)}{\Gamma\left(m/(q-1)\right)},
\end{eqnarray}
and the inverse transformation 
\begin{equation}
G_{3/2}(z,q)=\frac{1}{2\pi i}\int_{c-i\infty}^{c+i\infty}\frac{1}{(q-1)^{3/2+s}}\Gamma(s)
\sum_{m=1}\frac{\Gamma\left(m/(q-1)-(3/2)-s\right)}{\Gamma\left(m/(q-1)\right)}\alpha^{-s}ds,
\end{equation}
where the contour of integration closes on the left half plane.
With use of the approximation \cite{AS}
\begin{equation}
\frac{\Gamma\left(m/(q-1)-(3/2)-s\right)}{\Gamma\left(m/(q-1)\right)}=\frac{1}{(m/(q-1))^{3/2+s}}
\left[1+\frac{b}{m/(q-1)}+\frac{c}{(m/(q-1))^2}+...\right],
\end{equation}
with the constants
\begin{eqnarray}
b&=&\frac{1}{2}(3/2+s)(3/2+s+1),\nonumber\\
c&=&\frac{1}{12}\frac{\Gamma(-3/2-s+1)}{\Gamma(3)\Gamma(-3/2-s-1)}(3(5/2+s)^2+s+1/2),
\end{eqnarray}
the function $G_{3/2}(z,q)$ is written
\begin{eqnarray}
G_{3/2}(z,q)&=&\frac{1}{2\pi i}\int_{c-i\infty}^{c+i\infty}\frac{\Gamma(s)\alpha^{-s}}{(q-1)^{3/2+s}}
\left[(q-1)^{3/2+s}\zeta(3/2+s)+b(q-1)^{5/2+s}\zeta(5/2+s)\right.
\nonumber\\
&+& \left.c(q-1)^{7/2+s}\zeta(7/2+s)+...\right]ds.
\end{eqnarray}
The function $\Gamma(s)$ has simple poles at $s=-n$ with residues $(-1)^n/n!$ and the
zeta function $\zeta(w)$ has a simple pole at $w=1$ with residue equal to $+1$.
With use of the residue theorem the function $G_{3/2}(z,q)$ is expressed 
as a power series of $\alpha$ as follows
\begin{equation}
G_{3/2}(z,q)=\frac{1}{(q-1)^{3/2}}\sum_{m=1}\sum_{n=0}\frac{(-\alpha(q-1))^n}{n!}\frac{\Gamma(m/(q-1)-3/2+n)}
{\Gamma(m/(q-1))}+ \Gamma(-1/2)\alpha^{1/2}-\frac{1}{6}(q-1)^2\Gamma(-5/2)\alpha^{5/2}+O(\alpha^{7/2}).
\end{equation}
A simple check shows that the series with integer powers of $\alpha$ reduces, as $q\rightarrow 1$,
to the  result in Ref.\cite{R} $\sum_{n=0}(-\alpha)^n\zeta(3/2-n)/n!$. At lowest order in
$\alpha$ we obtain
\begin{equation}
G_{3/2}(z,q)\approx \frac{1}{(q-1)^{3/2}}\sum_{m=1}\frac{\Gamma(m/(q-1)-3/2)}{\Gamma(m/(q-1))}
+\Gamma(-1/2)\alpha^{1/2}.\label{G1}
\end{equation}
The series in equation (\ref{G1}) approximates at lowest order to the function $\zeta(3/2)$
and the last term is identical to the standard, $q=1$, result. Therefore, 
since for $\alpha\approx 0$ we have $G_{3/2}(z,q)\approx \alpha^{1/2}$, defining
$t=(T-T_c)/T_c$ we find that $\alpha\sim t^2$.   In addition, 
the correlation length $\xi\sim\lambda t^{-1}$,
as the standard case.  Since, near the
critical temperature,  the
$t$ dependence of the functions $G_{\nu}(z,q)$ is the same as the case of
the Bose-Einstein functions $g_{\nu}(z)$ and   the
thermodynamic functions of the two systems have the same functional form in terms of
$G_{\nu}$ and $g_{\nu}$, the remaining critical exponents are also independent of $q$.
\section{Conclusions}\label{3}
In this paper our main concern has been to study the long range
behavior predicted by thermodynamical systems described by a  factorized
partition function 
. As pointed out in the Introduction, this factorized
partition function has been shown to approximate well the partition
function of nonextensive statistics.  A calculation of the correlation
functions indicate that a Bose gas for the $q\neq 1$ case
is less correlated than for $q=1$.
 In particular, we showed  that the correlation length for $q=1$ is
larger than for $q\neq 1$ at all temperatures , and
the critical exponents are independent of $q$.  Certainly,  our calculations
can draw conclusions   on the long range 
behavior of a thermodynamical system in the factorization approach only. On the other hand, correlations in  nonextensive thermodynamics  
should be studied with use of the Tsallis' partition function. The fact that a Bose-Einstein
gas obeying the studied factorized partition function is less correlated
than for the Boltzmann-Gibbs case implies that the long range behavior
expected in nonextensive thermodynamics is lost when the Tsallis' partition function is replaced
by a factorized one.  Thus, our results point out that
the errors introduced by forcing factorization, which could be
not significant for the evaluation of thermodynamic functions, are large enough
 that the long range behavior
of Tsallis thermostatistics cannot be studied within this factorization approach.

\newpage
\epsfxsize=400pt \epsfbox{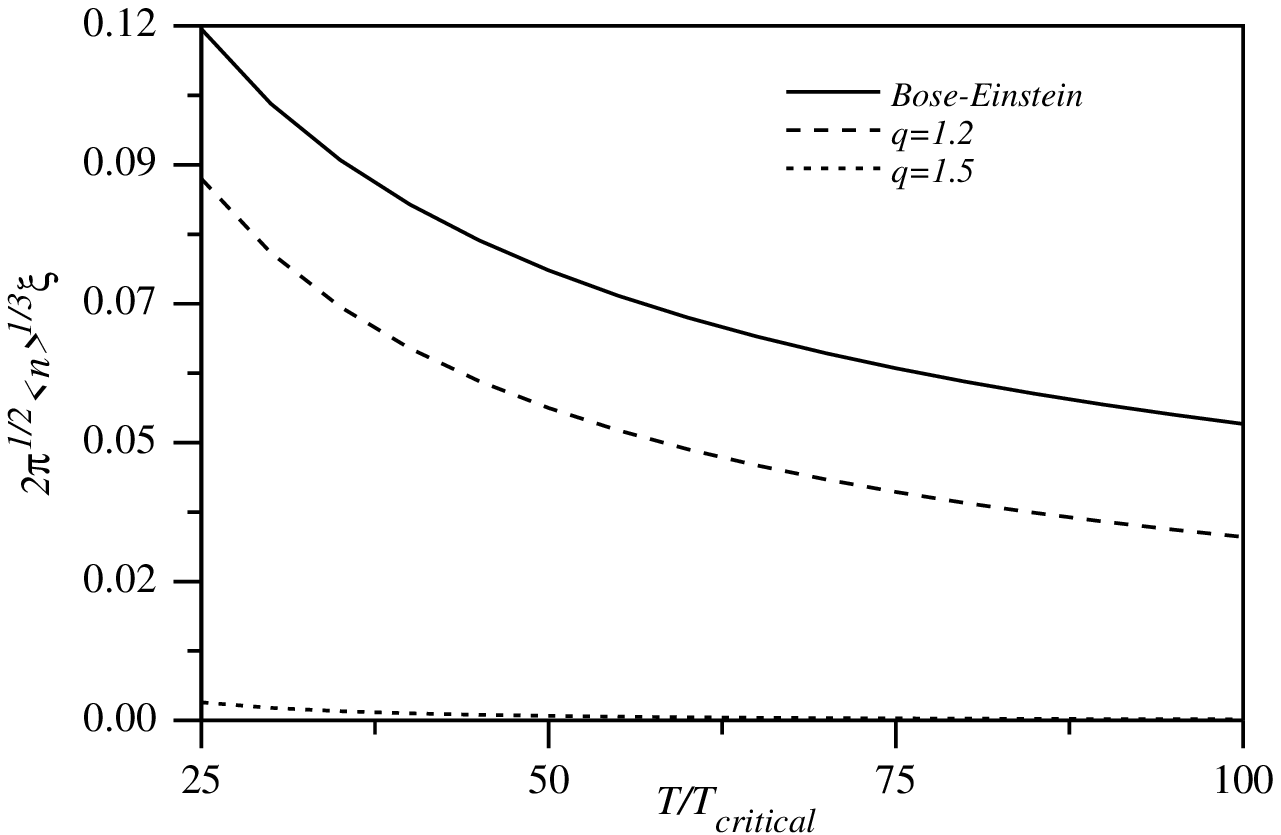}
\newpage
\epsfxsize=400pt \epsfbox{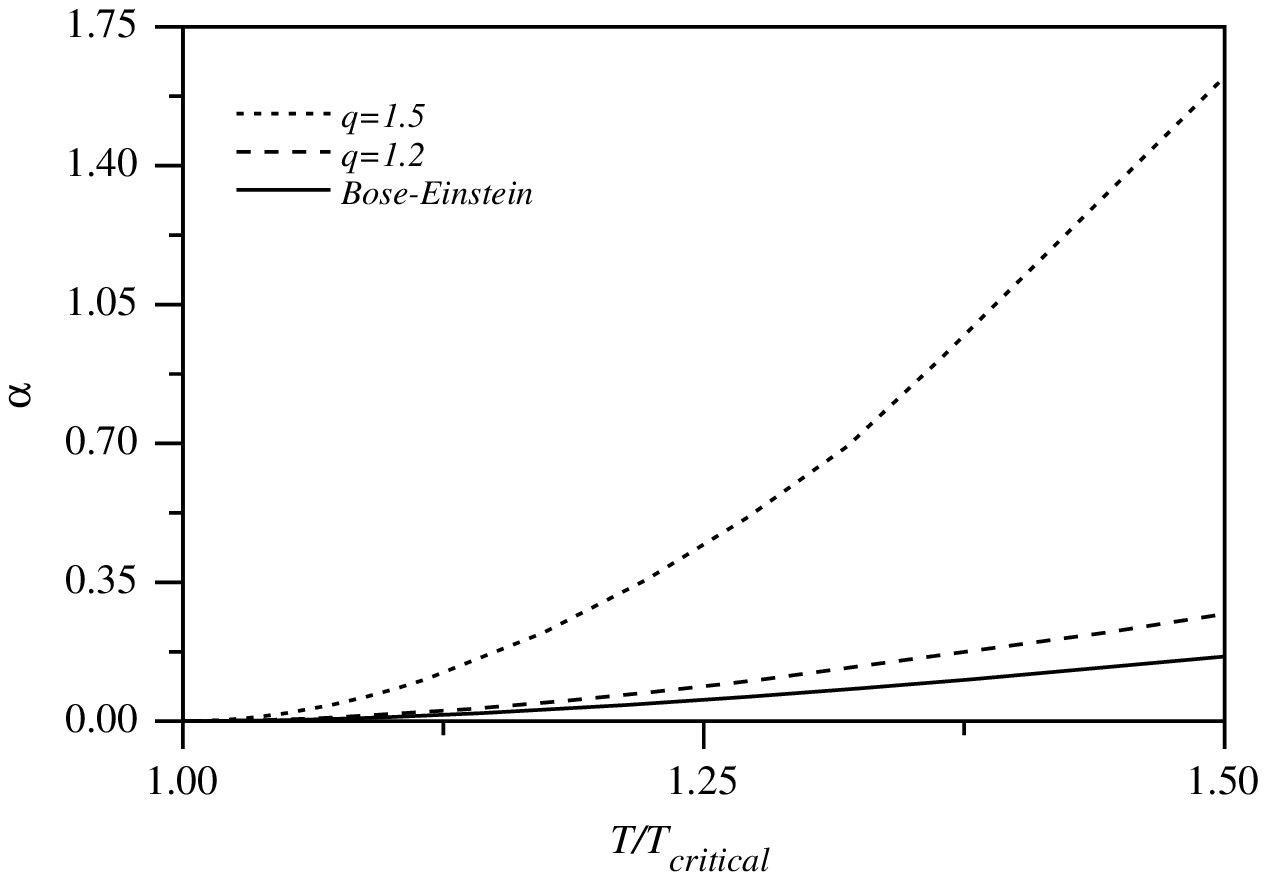}
\newpage
FIG. 1. The  function $f=2\pi^{1/2}\langle n\rangle^{1/3}\xi$, where 
$\langle n\rangle$ is the average number of particles per volume and
$\xi$ is the correlation length at high temperatures for
the cases of $q=1,6/5,3/2$ as a function of $T/T_{critical}$. 
$T_{critical}$ refers to the critical temperature for each value of $q$. This graph
shows that the correlation length decreases as the value of $q$ increases.\\
\\
\\
FIG. 2.  The behavior of  $\alpha=-\beta\mu$ as a function of the temperature for the cases
 $q=1,6/5,3/2$, indicating that for low temperatures a Bose-Einstein gas is 
more strongly correlated for $q=1$.

\end{document}